\documentclass[twocolumn,nofootinbib,amsmath,prl,aps,superscriptaddress,tightenlines,preprintnumbers]{revtex4}

\usepackage{graphicx}
\usepackage{braket}
\usepackage[usenames]{color}
\usepackage{latexsym}
\usepackage{bm}
\usepackage{mathrsfs}
\usepackage[caption=false]{subfig}

\usepackage{tikz}
\usetikzlibrary{arrows,decorations.pathmorphing,backgrounds,positioning,fit,petri,automata,shadows,calendar,mindmap,
decorations.markings,calc}

\usepackage[utf8]{inputenc}
\usepackage{graphicx}

\def\bea{\begin{eqnarray}}
\def\eea{\end{eqnarray}}

\begin{document}

\newcount\hour \newcount\minute
\hour=\time \divide \hour by 60
\minute=\time
\count99=\hour \multiply \count99 by -60 \advance \minute by \count99
\newcommand{\mydate}{\ \today \ - \number\hour :00}

\title{Interpreting $W$ mass measurements in the SMEFT.}

\author{Mikkel Bj\o rn and Michael Trott,\\
Niels Bohr International Academy,
University of Copenhagen, \\
Blegdamsvej 17, DK-2100 Copenhagen, Denmark
}

\begin{abstract}
Measurements of the $W^\pm$ mass ($m_W$) provide an important consistency check of the Standard Model (SM) and
constrain the possibility of physics beyond the SM. 
Precision measurements of $m_W$ at hadron colliders are inferred from kinematic distributions of transverse variables. 
We examine how this inference is modified when considering the presence of physics beyond the SM expressed in terms of local contact operators.
We show that Tevatron measurements of $m_W$ using transverse variables are transparent
and applicable as consistent constraints in the Standard Model Effective Field Theory (SMEFT) with small measurement bias. This means that the leading challenge to interpreting these measurements
in the SMEFT is the pure theoretical uncertainty in how these measurements are mapped to Lagrangian parameters.
We stress the need to avoid using naive combinations of Tevatron and LEPII 
measurements of $m_W$ without the introduction of any SMEFT theoretical error to avoid implicit UV assumptions. \end{abstract}

\maketitle
\newpage

\paragraph{\bf I. Introduction.}
The lack of any statistically significant deviation from the Standard Model (SM) to date, argues that a mass gap is present
between the scale of new physics -- $\Lambda$ -- and the electroweak scale $v \simeq 246 \, {\rm GeV}$. Such a mass gap, if limited to $v/\Lambda \lesssim 1/4 \, \pi$ can also be consistent with expectations of UV physics motivated by naturalness concerns for the Higgs mass ($m_h$). Broad classes of new physics scenarios consistent with this assumption can be studied efficiently using Effective Field Theory (EFT) methods to analyse data limited to energies $\sqrt{s} \sim v << \Lambda$. This includes future data gathered on the ``Higgs pole" where $\sqrt{s} \sim m_h$.

 It is of interest to examine possible future deviations in such measurements in a consistent theoretical framework that also incorporates lower 
energy experimental data gathered on and near the $Z$ and $W^\pm$ poles. This theoretical framework
has come to be known as the Standard Model 
Effective Field Theory (SMEFT). Near $Z$ pole data can be directly incorporated in the SMEFT \cite{Grinstein:1991cd,Han:2004az,Pomarol:2013zra,Ciuchini:2014dea,Ellis:2014jta,Petrov:2015jea,Berthier:2015oma,Berthier:2015gja,David:2015waa} by interpreting the well known LEP electroweak pseudo-observables \cite{ALEPH:2005ab} in terms of constraints
on higher dimensional operators. Many operators at dimension six can contribute to these observables, and further measurements are required to constrain all the operators that can contribute to such LEP data without flat directions
in the SMEFT parameter space. Measurements of the $W^\pm$ boson's decay width and mass ($\bar{\Gamma}_W, \bar{m}_W$) are a particularly  important further source of constraints on the SMEFT in this sense. Note that we use bar superscripts to denote the canonically normalized parameters in the SMEFT, and hat superscripts to denote parameters derived from the measurements of our input parameter set $(\hat G_F, \hat m_Z, \hat \alpha)$ at tree-level in the SM. Measurements of these parameters are more challenging to incorporate consistently in global studies, as the leptonic decay of the $W^\pm$ blocks precise direct experimental reconstruction of a mass peak due to the presence of final state missing energy. This challenge is overcome with the use of transverse kinematic variables, where a Jacobian peak in the corresponding distributions are exploited to extract $\hat{m}_W^2$ \cite{vanNeerven:1982mz,Arnison:1983rp,Banner:1983jy,Smith:1983aa,Barger:1983wf}. Modern fits to $\hat{m}_W^2$ reported using this technique with Tevatron data obtain the values in Table 1. 
\begin{figure}
		\includegraphics[width=0.15\textwidth]{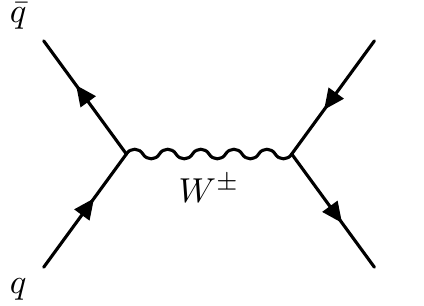}
		\includegraphics[width=0.15\textwidth]{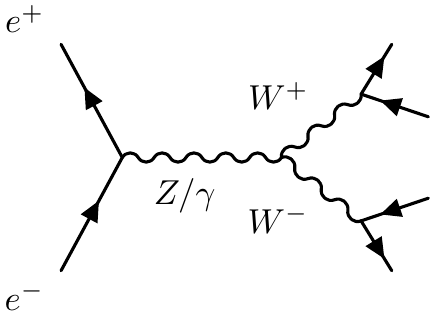}
		\includegraphics[width=0.15\textwidth]{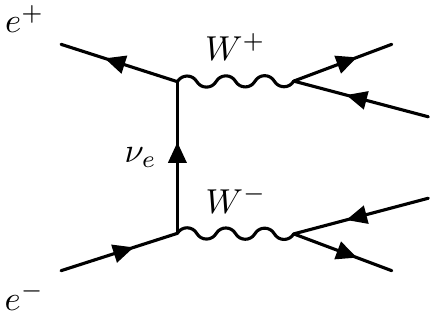}
	\caption{Left diagram is the leading order $W^\pm$ production process at the Tevatron. The remaining figures show
	the leading order CC03 processes (defined in a double pole approximation) used to extract $\hat{m}_W$ at LEPII.}
	\label{fig:phase_space}
\end{figure}
Determinations of $\hat{m}_W$ at LEPII use two distinct methodologies. In one of these, scans of the $d \sigma/d \sqrt{s}$ differential cross section in the threshold region ($\sqrt{s} \simeq 2 \, m_W$) exploit the rise in the cross section proportional to the velocity of the $W^\pm$ bosons, given by $\beta = \sqrt{1- 4 \bar{m}_W^2/s}$, to extract $\hat{m}_W$.
This is done for the processes $e^+ e^- \rightarrow W^+ \, W^- \rightarrow \bar{f} \, f \,\bar{f} \, f$, with $f = \{\ell,q\}$ at LEPII, and this approach is relatively insensitive to some corrections \cite{Hagiwara:1987pb} that could come about in the SMEFT. Such corrections include modifications of the couplings of the $W^\pm$ and $Z$ to the initial and final state fermions, and possible anomalous Triple Gauge Coupling (TGC) parameters. The presence of these corrections is a related concern when interpreting the results of the second measurement employed at LEPII, where $\hat{m}_W$ is extracted via kinematic fits to the W pair invariant mass distribution away from the threshold region. 
This issue is related to the issue we study in detail at the Tevatron, and we return to this point in Section VI. We stress that all the experimental approaches reported in Table 1 are appropriate when
consistency testing the SM, and can be combined as in Ref. \cite{Agashe:2014kda,Schael:2013ita,Group:2012gb} under the SM assumption. 

This paper is aimed at the consistent interpretation of $\hat{m}_W$ measurements in a different field theory than the SM, the SMEFT. When all dimension six operators are allowed to be present with arbitrary Wilson coefficients, interpretation of the measurements is modified. This modification comes about by an impact on the measurement itself, and in the mapping of the experimental results to different
Lagrangian parameters. In this paper, we focus on the first question. We examine whether a bias on the measured $\hat{m}_W$ arises due to SMEFT corrections and thus whether an additional theoretical error of this form should be included in fits of SMEFT parameters due to such a bias. After describing the modifications of $W^\pm$ parameters in the SMEFT in Section II, the potential bias is analysed in detail for the Tevatron measurement in Sections III-V. In Section VI we comment on the LEPII measurements, whereas Section VII is reserved for conclusions


 \begin{center}
\begin{table}[t]
\centering
\tabcolsep 8pt
\begin{tabular}{|c|c|c|}
\hline 
Result & Value & Ref.  \\ \hline
{\rm D\O } & 80.375 $\pm$ 0.023 &\cite{Abazov:2012bv} \\
{\rm CDF} & 80.387 $\pm$ 0.019 &\cite{Aaltonen:2012bp} \\
\hline
{\rm Tev. Comb.} & 80.387 $\pm$ 0.016 &\cite{Agashe:2014kda} \\
\hline
{\rm LEP threshold} & 80.42 $\pm$ 0.20 $\pm$ 0.03 &\cite{Schael:2013ita} \\
{\rm LEP direct} & 80.375 $\pm$ 0.025 $\pm$ 0.022 & \cite{Schael:2013ita}  \\
\hline
{\rm LEP. Comb.} & 80.376 $\pm$ 0.033 &\cite{Schael:2013ita} \\
\hline \hline
{\rm Global Comb.} & 80.385 $\pm$ 0.015 &\cite{Group:2012gb} \\
\hline
\hline
\end{tabular}
\caption{$W^\pm$ mass measurements reported by the Tevatron and LEPII collaborations.}
\end{table}
\vspace{-0.8cm}
\end{center} 
\paragraph{\bf II. The $W^\pm$ mass and Width in the SMEFT.}
In the SM, at tree level, the input parameter set $\{\hat{m}_Z,\hat{G}_F,\hat{\alpha}_{ew}\}$ fix  
$\hat{m}_W^2 = 2 \, \sqrt{2} \, \pi \hat{\alpha}_{ew}/(\hat{G}_F \, s_{\hat{\theta}}^2 )$ where
\bea
s_{\hat{\theta}}^2 = 1/2 - \sqrt{1- 4 \, \pi \hat{\alpha}_{ew}/\sqrt{2} \hat{G}_F \hat{m}_Z^2}.
\eea
The shift in the pole mass $\delta m_W^2 = \hat{m}_W^2- \bar{m}_W^2$ in the $\rm{U}(3)^5$ symmetric version of the SMEFT is given by \cite{Berthier:2015oma}
\bea\label{mwshift}
\frac{\delta m_W^2}{\hat{m}_W^2} = \hat{\Delta} \, \left[4  \, C_{HWB} + \frac{c_{\hat{\theta}}}{s_{\hat{\theta}}} C_{HD} + 4 \frac{s_{\hat{\theta}}}{c_{\hat{\theta}}} C_{H \ell}^{(3)} - 2 \frac{s_{\hat{\theta}}}{c_{\hat{\theta}}} C_{\ell \, \ell}\right],
\eea
where $\hat{\Delta} =  c_{\hat{\theta}} s_{\hat{\theta}} /(c^2_{\hat{\theta}}-s^2_{\hat{\theta}}) \, 2 \sqrt{2} \hat{G}_F$. We use a $\delta$ to indicate a shift in a quantity due to the complete set of corrections present at leading order in the power counting in the SMEFT. We use the Warsaw basis
of dimension six operators in the SMEFT \cite{Grzadkowski:2010es} that defines the Wilson coefficients $C_i = \{C_{HWB},C_{HD},C_{H \ell}^{(3)},C_{\ell \, \ell} \}$. 
The cut off scale has been absorbed into the definition of the Wilson coefficients so that the mass dimension of these parameters is $-2$.
The value of $\bar{m}_W^2$ can be predicted in the SM with complete one loop \cite{Denner:1990tx}  and even full two loop corrections
\cite{Awramik:2003rn}. The full one loop corrections
compared to the Born approximation to $\bar{\Gamma}_W$ are a $\sim 2 \%$ \cite{Beenakker:1996kt} correction. The size of this correction depends on how the
tree level value of  $\bar{\Gamma}_W$ is related to the input observables. Absorbing universal radiative corrections into
the parameters defining the width in an improved Born approximation reduces the size of the remaining perturbative corrections at one loop to $\sim 0.5 \%$
\cite{Rosner:1993rj,Beenakker:1996kt,Denner:1990tx}. The effect of still neglected higher order perturbative terms is then expected to be a loop factor smaller than this variation. 

The expression for the shift of $\Gamma_W$ in terms of the input parameter $\hat{G}_F$ and the derived value $\hat{m}_W$ 
in the SMEFT is given by
\bea\label{deltam}
\frac{\delta \Gamma_W}{\hat{\Gamma}_W} = \frac{\sqrt{2}}{3 \, \hat{G}_F} \left[C_{H \ell}^{(3)}+ 2 \, C_{H q}^{(3)}\right] - \sqrt{2} \, \delta \hat{G}_F -  \frac{3}{2} \, \frac{\delta m_W^2}{\hat{m}_W^2}. 
\eea
where $\hat{\Gamma}_W = 9 \sqrt{2} \hat{G}_F \hat{m}_W^3/(12 \, \pi)$. 
Here  $\delta \hat{G}_F  = C_{H \ell}^{(3)}/\hat{G}_F - C_{\ell \, \ell}/2 \hat{G}_F$ and we are considering the massless fermion limit.  Being conservative, so long as $\delta \Gamma_W/\bar{\Gamma}_W \gtrsim 0.5 \%$ it is clear that neglected
SMEFT corrections can have a non negligible impact on the theory error of an extracted value of $\hat{m}_W$ at the Tevatron.
This condition corresponds to a bound on the Wilson coefficients and the cut off scale of the form $ \Lambda/\sqrt{C}_i \lesssim 3.5 \, {\rm TeV}$, which are the cases of interest motivated by the hierarchy problem.

\paragraph{\bf III. Spectra for Extractions of $\bm{{m}_W}$ at the Tevatron.}
We illustrate
the effect of generalizing these measurements into the SMEFT following the analytic methods of Ref.\cite{vanNeerven:1982mz}. This is sufficient for our purposes as detector resolution effects can only be approximated without direct access to the experimental data, and are substantial. 

The value of $\bar{m}_W^2$ is extracted from Tevatron data using kinematic templates for distributions in the variables $m_T,P_{T\ell}, E_T \! \! \! \! \! \! / \, \, \,$.
The latter is found to have a small effect on the fit \cite{Abazov:2012bv}, so we neglect this spectrum. We define the transverse mass to be 
$m_T^2 = 2 P_{T,\ell} \, P_{T,\nu} (1- \cos \, \theta_{\ell \nu})$, with $P_{T,\ell/\nu}$ and $\theta_{\ell \nu}$ the momenta and angle between the leptons in the perpendicular plane 
to the $\bar{p} \, p$ collision axis. We agree with the result in Ref.\cite{vanNeerven:1982mz} for  $d \sigma/d m_T$  for the effective kinematic spectra once the partonic production mechanism  is factorized out, consistent with a narrow width expansion, and integrated over Parton Distribution Functions (PDF's), generating an effective $p_T$ for $W^\pm$. Note however the correction to the last term in the numerator of the $I$ function (derived from the Jacobian), correcting a typo in Ref.\cite{vanNeerven:1982mz}. Here $\mu^2 = m_T^2/s'$, $\alpha =(\gamma^2 -1)^{1/2} = P_{T,W}/\sqrt{s'}$ and $\gamma = \sqrt{P_{T,W}^2 + s'}/\sqrt{s'}$. $P_{T,W}$ is the transverse momentum of the $W^\pm$ boson present due to the effects of PDFs.
We find the following results for $d \sigma/d m_T$
\begin{widetext}
\bea
\frac{d \, \sigma}{d m_T} &=& \frac{1}{(\pi/2 + \arctan \bar{m}_W/\bar{\Gamma}_W)} \int_{m_T^2}^\infty  \! \! \! \! d s' \, \frac{ \bar{m}_W \, \bar{\Gamma}_W}{(s' - \bar{m}_W^2)^2 + \bar{m}_W^2 \, \bar{\Gamma}_W^2} \frac{m_T}{(s'(s'-m_T^2))^{1/2}} \, \int_0^{2 \pi}  \! \! \! \! d \phi \sum_{ij} \frac{d \sigma_{ij}}{d \cos \theta} \, I(\mu, \phi,\alpha), \nonumber \\
I(\mu, \phi,\alpha) &=& \frac{\mu^4 + \mu^4 \alpha^2 \cos^2 \phi + 2 \mu^2 \alpha^2 \sin^2 \phi + \alpha^4 \sin^2 \phi}{(\mu^2+ \alpha^2 \sin^2 \phi)^{1/2} (\mu^2+\mu^2 \alpha^2 \cos^2 \phi + \alpha^2 \sin^2 \phi)^{3/2}}.
\eea
\end{widetext}
Still following Ref.\cite{vanNeerven:1982mz}, we have introduced a normalized Breit-Wigner resonance, consistent with a narrow width normalization in the limit $\bar{\Gamma}_W/\bar{m}_W \rightarrow 0$ for the unstable $W^\pm$ boson. The angular dependence and normalization of the partonic $\sigma(\bar{q}_i \, q_j \rightarrow W^\pm \rightarrow \ell^{\pm} \, \nu)$ is given by
\begin{align}\label{eq:original_dist}
 \frac{d \sigma_{ij}}{d \cos \theta} &=& \tilde \sigma_{ij}\frac{3 \, (\hat{G}_F \hat{M}_W^2)^2 |V_{ij}|^2}{8 \, \sqrt{2} \pi \, N_c \, s} |\bar{g}^{W,q}_{ij}|^2 \,  \left[1+ \cos^2 \theta\right] {\rm Br}
\end{align}
where $ {\rm Br} = \sum_i {\rm Br} (W^\pm \rightarrow \ell_i^\pm \nu)$ is the sum of the branching ratios to the specific lepton final states included in the analysis, and $\tilde \sigma_{ij}$ is a PDF dependent production cross section factor, that can be taken to be constant within $O(\bar \Gamma_W/\bar m_W)$, since the main SMEFT dependence of the production cross section has been included, by pulling $|\bar{g}_W^q|^2$ out of this production cross section. We follow Ref.~\cite{Berthier:2015oma} absorbing the SMEFT shifts of the input parameters into the redefined $W^\pm$-coupling $\bar g^{W,\ell / q}$. 
We have also included the initial state partonic factor of $1/s$ and note that in our numerical simulations this leading factor of $1/s$ is replaced with $1/\hat{m}_W^2$ consistent with the narrow width approximation factorization of the process.
The angular dependence here is defined in the un-boosted $W^\pm$ rest frame  w.r.t a $z$ axis along the $\bar{p}$ direction, with the electron and neutrino momentum decomposed as
\bea
p_\ell &=& (\sqrt{s'}/2) \{1, \sin \theta \, \cos \phi,\sin \theta \, \sin \phi, \cos \theta \}, \\
p_\nu &=& (\sqrt{s'}/2) \{1, -\sin \theta \, \cos \phi, -\sin \theta \, \sin \phi, - \cos \theta \}.
\eea
We express $\cos \theta$ in terms of $\mu$
\begin{align}
    \cos^2 \theta &= \frac{(1-\mu^2)(\mu^2 + \alpha^2 \sin^2 \phi)}{\mu^2(1+\alpha^2 \cos^2 \phi)+\alpha^2 \sin^2 \phi} \\
    &= \gamma^2 \left(1-\frac{P_-^2}{\gamma^2 s'}\right)\left(1-\frac{P_+^2}{\gamma^2 s'}\right)
\end{align}
where, having denoted $P_\pm = (P_{T,\ell} \pm P_{T,\nu})$, we have also related $\theta$ to the lepton transverse momenta.
A similar calculation gives the result for the $P_T^\ell$ spectrum 
\begin{widetext}
\bea \label{eq:pt_spectrum}
\frac{d \, \sigma}{d P_{T,\ell}}  &=& \frac{1}{(\pi/2 + \arctan \bar{m}_W/\bar{\Gamma}_W)} \int_{R_s}^\infty  \! \! \! \! d s' \, \frac{ \bar{m}_W \, \bar{\Gamma}_W}{(s' - \bar{m}_W^2)^2 + \bar{m}_W^2 \, \bar{\Gamma}_W^2}   \int_{R_a}^{R_b}  \! \! \! \! d P_{T,\nu}  \sum_{ij} \, \frac{d^2 \sigma_{ij}}{d P_{T,\ell}\;d P_{T,\nu}}, 
\\ \label{eq:ptlptn_distribution}
 \frac{d^2 \sigma_{ij}}{d P_{T,\ell}\;d P_{T,\nu}} &=& \frac{6 \, (\hat{G}_F \hat{M}_W^2)^2 |V_{ij}|^2}{\sqrt{2} \pi \, N_c \, s} |\bar{g}^{W,q}_{ij}|^2 \, {\rm Br}
  \frac{P_{T,\ell} \, P_{T,\nu}
  [1+\gamma^2(1-P_-^2/\gamma^2s')(1-P_+^2/\gamma^2s')]}{\sqrt{\gamma^2s'-P_+^2}\sqrt{\gamma^2s'-P_-^2}\sqrt{P_+^2-\alpha^2s'}\sqrt{\alpha^2s'-P_-^2}}.
\eea
\end{widetext}
We find the phase space in this case to be
\begin{subequations}\begin{align}
	R_s &= \left \lbrace \begin{array}{l} 0 \\ 
    4 P_{T,\ell}(P_{T,\ell}-\alpha \sqrt{s'}) \end{array} \right. &&\left.\begin{array}{l} P_{T,\ell} < \alpha \sqrt{s'}, \\
     P_{T,\ell}>\alpha \sqrt{s'},\end{array} \right. \nonumber \\
    R_a &= \left \lbrace \begin{array}{l}
     \alpha \sqrt{s'}-P_{T,\ell} \\  P_{T,\ell}-\alpha \sqrt{s'}
    \end{array} \right. &&\left.\begin{array}{l} 
    P_{T,\ell} < \alpha \sqrt{s'}, \\ P_{T,\ell}>\alpha \sqrt{s'},\end{array} \right. \nonumber \\
    R_b &= \left \lbrace \begin{array}{l}
     \alpha \sqrt{s'}+P_{T,\ell} \\  \gamma \sqrt{s'}-P_{T,\ell}
    \end{array} \right. &&\left.\begin{array}{l} 
    P_{T,\ell} < (\gamma-\alpha)\sqrt{s'}/2, \\ P_{T,\ell} > (\gamma-\alpha)\sqrt{s'}/2.\end{array} \right. \nonumber
\end{align}\end{subequations}

\paragraph{\bf IV. Variation in Extractions of $\bm{{m}_W}$.}
Kinematic template fits to extract $\hat{m}_W$ can be impacted by the presence of local contact operators in the SMEFT as follows.
The pole mass $\bar{m}_W^2$ is shifted compared to the expected value in the SM, as given in Eq. (2).
This is how the constraint on $\hat{m}_W$ measurements is intended to impact the fit constraint space
of the SMEFT, at leading order in $v_T^2/\Lambda^2$. The value of the width is modified, see Eq. (3). This shift is assumed to vanish when SM extractions
of $\hat{m}_W^2$ are performed and gives a theoretical error when interpreting these
measurements in the SMEFT. 

The overall normalization of both of the spectra is modified with a shift
\bea
\frac{\delta N_{ij}}{N_{ij}} =2 \left[ \frac{\delta g^{W,q}_{ij}}{V_{ij} \, \hat{g}^{W,q}}+ \frac{\delta g^{W,\ell}}{\hat{g}^{W,\ell}}\right] + \frac{1}{2} \, \frac{\delta m_W^2}{\hat{m}_W^2} - \frac{\delta \Gamma_W}{\hat{\Gamma}_W}.
\eea 
due to the SMEFT shift of the normalization of Eq. \eqref{eq:original_dist}.
This normalization effect is the correction that is not directly related to the $(\bar M_W,\bar \Gamma_W)$-dependent normalization of the Breit-Wigner function.

In a $\rm{U}(3)^5$ flavour symmetric scenario, we can relate the CKM matrix in the SMEFT to the CKM matrix in the SM in a straightforward fashion as
$\delta g^{W,q}_{ij} \propto V_{ij}$. The production spectra are the direct sum over all partonic quarks when considering hadronic collisions.
Incorporating PDF's, the partonic differential cross sections are convoluted and varied over the
PDF's in their $1 \sigma$ uncertainty band in the experimental analyses. The flavour dependence of $(\delta N)_{ij}$ is expected to be subdominant to this SM variation.
However, the remaining flavour universal variation due to $(\delta N)_{ii}$ is neglected in SM analyses, which leads to a further theoretical error.

Both $\delta N(=\sum \delta N_{ii})$ and $\delta \Gamma_W$ stem from shifts of the $W^\pm$-couplings, and are therefore correlated. We decompose
\begin{align}\label{eq:width_shift_decomp}
    \frac{\delta \Gamma_W}{\hat{\Gamma}_W} = \frac{\delta \Gamma_{||}}{\hat{\Gamma}_W} + \frac{\delta \Gamma_\perp}{\hat{\Gamma}_W},
\end{align}
where $\delta \Gamma_{||}$ captures all the correlation, and $\delta \Gamma_\perp$ corresponds to directions in Wilson-coefficient-space where the overall normalization of the spectra is unchanged. In the ${\rm U}(3)^5$ limit we find
\begin{align} \label{eq:decomp_coefficient}
    \frac{\delta N}{N}=\left(\frac{154 + 44 \, s_{\hat{\theta}}^2 - 38 s_{\hat{\theta}}^4}{227 +94 \, s_{\hat{\theta}}^2 -97 s_{\hat{\theta}}^4} \right)\frac{\delta \Gamma_{||}}{\hat{\Gamma}_W}
    \simeq 0.67 \frac{\delta \Gamma_{||}}{\hat{\Gamma}_W},
\end{align} where the proportionality factor is obtained by expressing $\delta \Gamma_W$ in a basis of the space spanned by $C_i$ that includes $\delta N/N$ as a basis vector. The relation is modified in a non-flavour-symmetric scenario, but the decomposition in Eq.\eqref{eq:width_shift_decomp} is still possible.

\paragraph{\bf V. Numerics.}
To estimate these effects, we generate fit templates for each spectra varying $\bar{m}_W^2$ in steps of 1.25 MeV around the central value of $\hat{m}_W^0=\hat{m}_W^{PDG}=80.385$ GeV \cite{Agashe:2014kda}. In generating these, we employ the tree level SM relation $\bar{\Gamma}_W \propto \bar{m}_W^3$, fixed so that $\hat{\Gamma}_W(\hat{m}_W^0)= \hat{\Gamma}_W^0 = \hat{\Gamma}_W^{PDG}=2.085$ GeV \cite{Agashe:2014kda}. We extract the $P_{T,W}$-distribution from the $P_{T,Z}$-distribution of Fig.(52) in \cite{D0:2013jba}, which is well approximated by the distribution \begin{align} \label{eq:ptw_dist}
    f(p_T) \propto \exp\left[\frac{-(p_T - p_0)^2}{2\, \sigma_0^2 \, p_T}\right]
\end{align} with mode $p_0 = 3.5$ GeV and $\sigma^2_0=3.15$ GeV. The $P_{T,W}$ follows the same distribution at a momentum scale lower by a factor of $m_W/m_Z$ \cite{privateD0}.
\begin{figure}
	\centering
	\includegraphics[width=0.9\columnwidth]{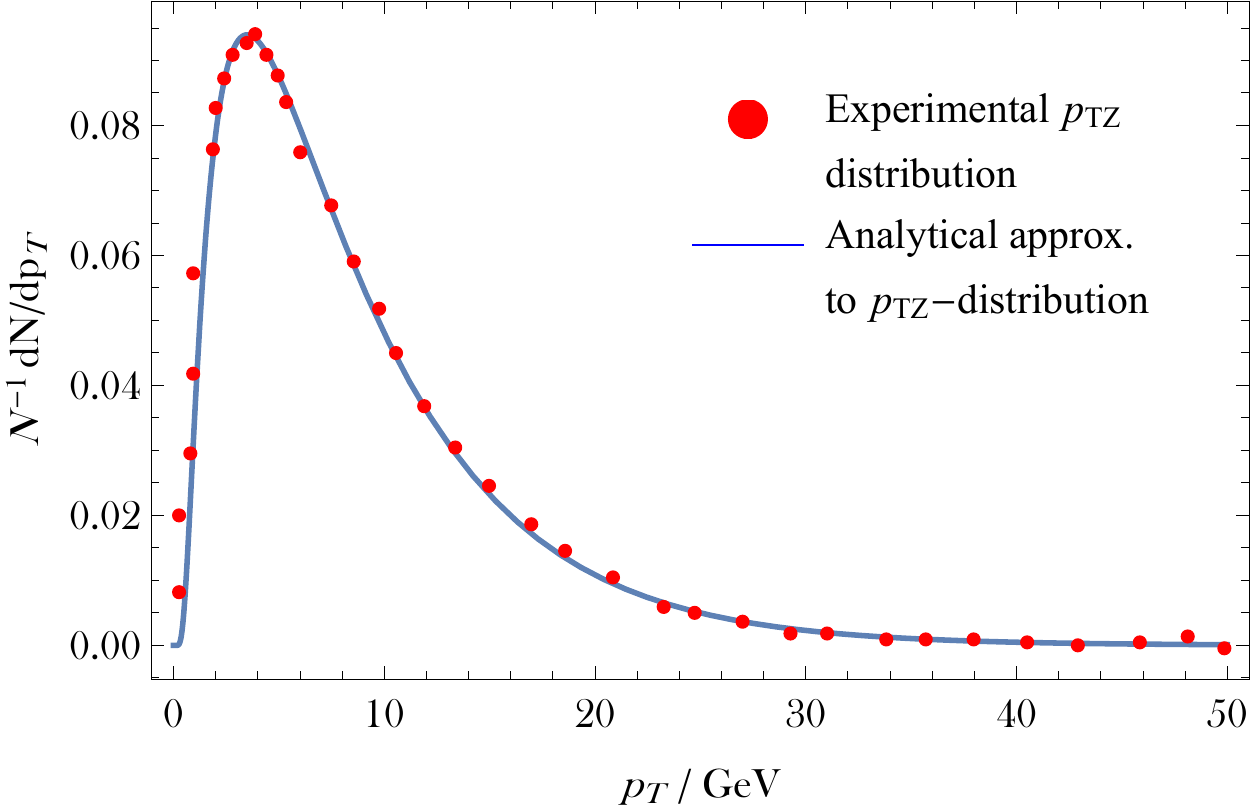}
	\caption{The experimental distribution of $p_{TZ}$ in $p\bar p \rightarrow Z \rightarrow e\bar e$ at the $D\emptyset$  detector, taken from Fig. (52) of \cite{D0:2013jba}, as well as the analytical approximation in \eqref{eq:ptw_dist}.}
	\label{fig:ptw_distribution}
\end{figure}
 The $m_T$-distribution is robust to variations of $p_{T,W}$ and we found it sufficient to set $p_{T,W}=p_{o,W}=(m_W/m_Z) \, 3.5$ GeV when generating the $m_T$-templates. The $p_{T,\ell}$-distribution is very dependent on $p_{T,W}$, and the result in Eq.\eqref{eq:pt_spectrum} was averaged over the distribution in Eq.\eqref{eq:ptw_dist} when generating $p_{T,\ell}$-templates. Detector effects are significant and we approximate them as a convolution of the calculated spectra with a Gaussian resolution function $R(x) = \mathcal{N}(0,\sigma)$, taking $\sigma \propto m_T/p_{T\ell}$. We choose the parameters to match the spectra published in Ref. \cite{Abazov:2012bv,D0:2013jba,Aaltonen:2012bp,Aaltonen:2013vwa} both with and without detector effects. This is an approximation to the true acceptance, which depends on the detailed kinematics and not just the variable $m_T, P_{T,\ell}$. However, it can reproduce the spectra of the above references reasonably well as shown in Fig. (\ref{fig:spectra}). 
 \begin{figure*}
\centering
\subfloat[]{
\includegraphics[width=0.45\textwidth]{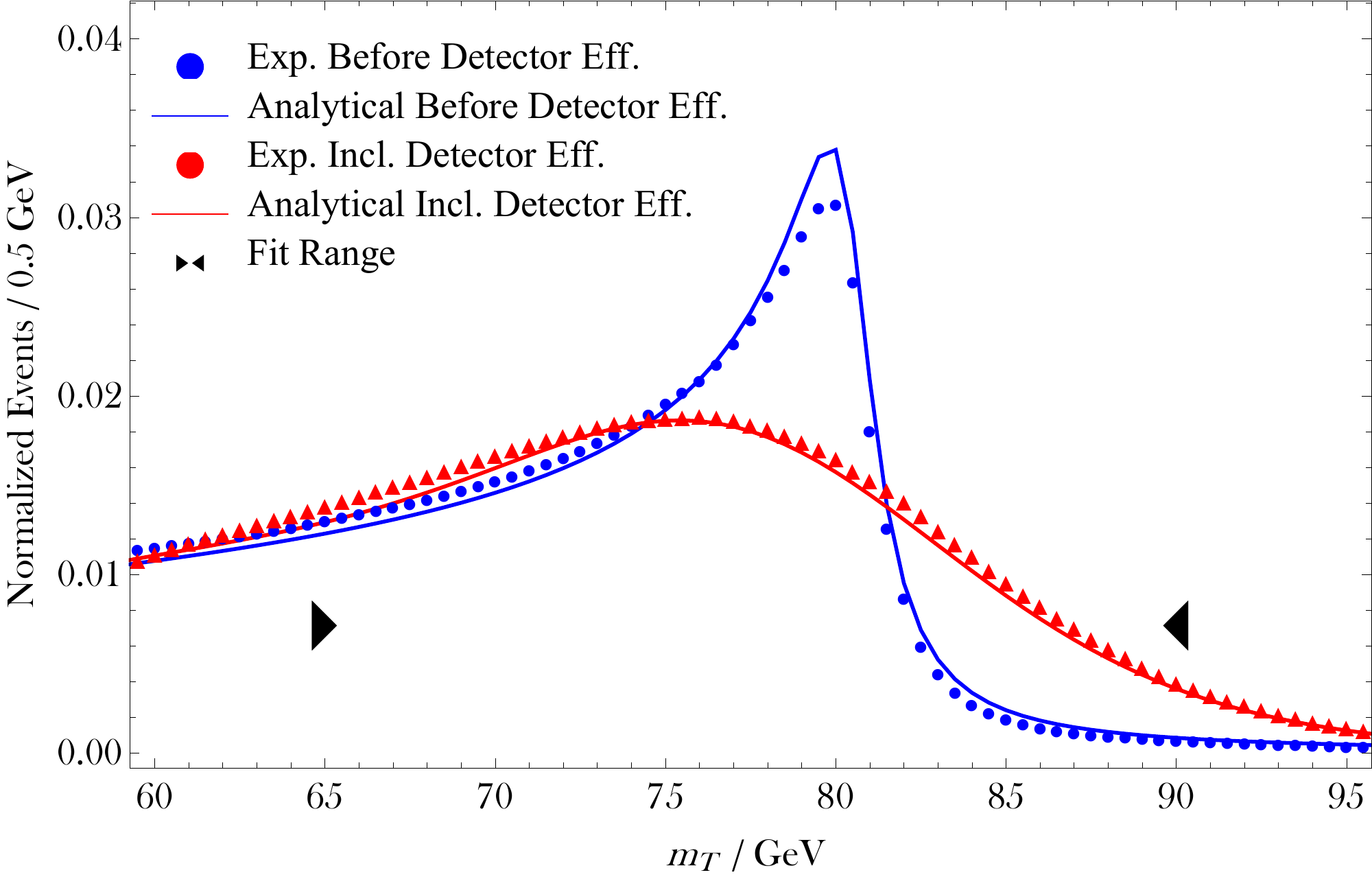}}
\hspace{0.1cm}
\subfloat[]{
\includegraphics[width=0.45\textwidth]{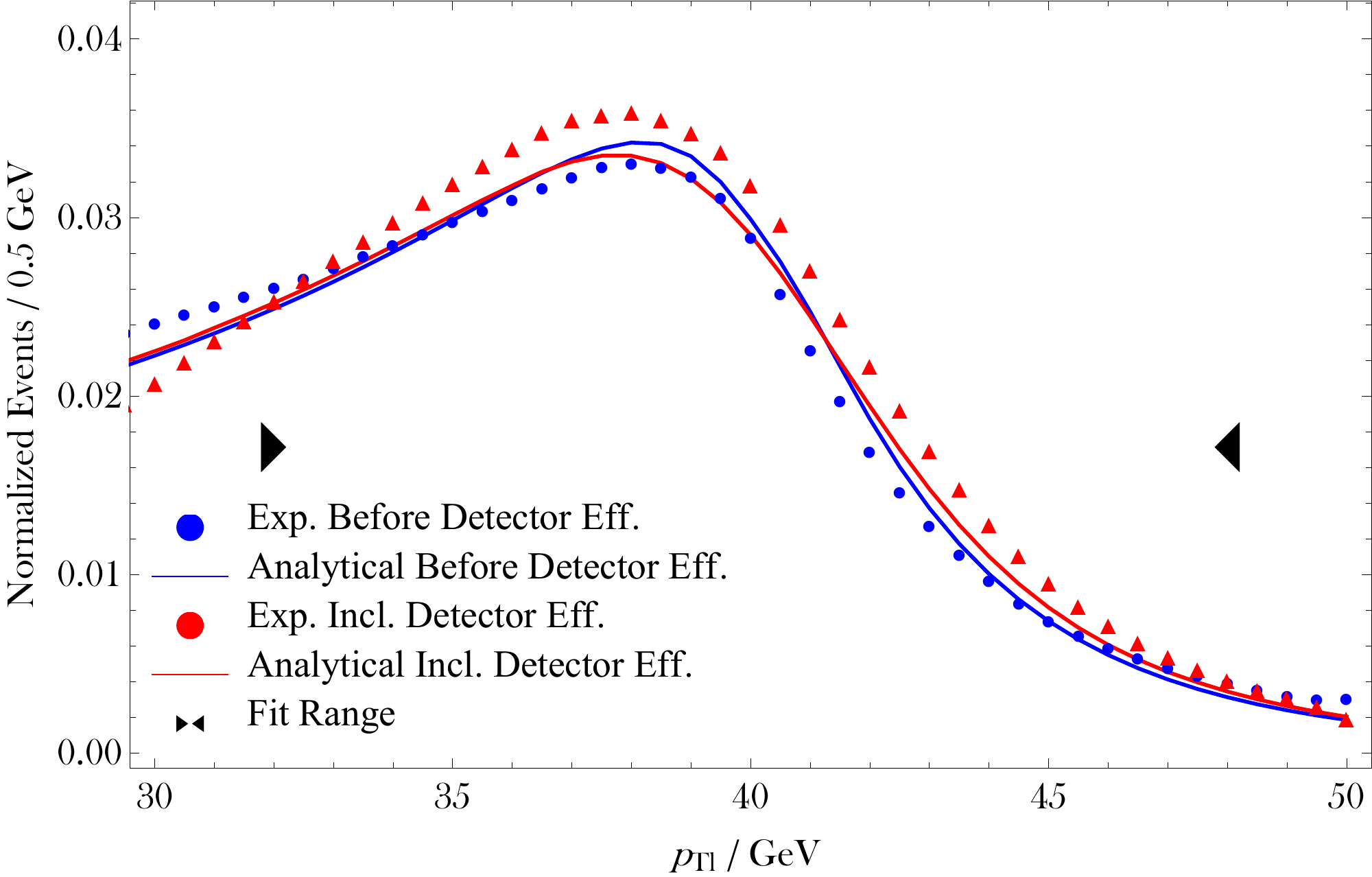}}
\caption{Comparison of generated (a) $m_T$-  and (b) $p_{T\ell}$-spectra with those observed in the $D\emptyset$ experiment \cite{D0:2013jba}. The corresponding figures for comparison with CDF \cite{Aaltonen:2013vwa} show the same degree of correspondence. \label{fig:spectra}}
\end{figure*} 
We choose an overall normalization to match the event counts, and thereby statistical significance found in the experiments,  by matching the peak-height of our spectra to the results in Ref. \cite{D0:2013jba,Aaltonen:2013vwa}, after approximating detector effects (thereby we eliminate the need to know $\tilde \sigma_{ij}$ and the integrated luminosity).

\begin{figure*}
\centering
\includegraphics[width=0.49\textwidth]{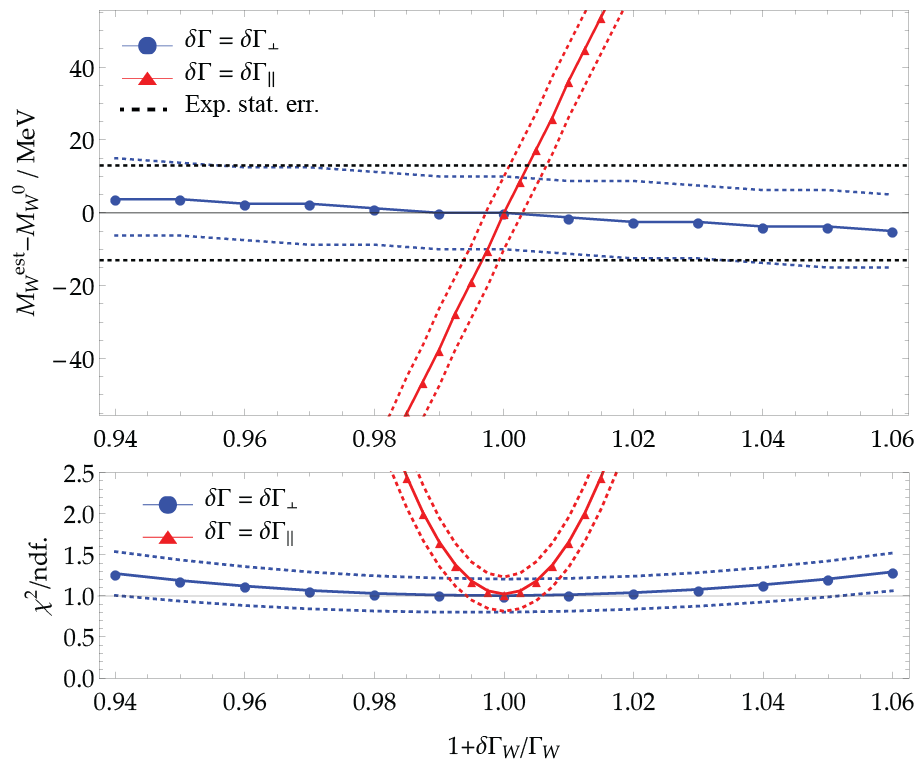}
\includegraphics[width=0.49\textwidth]{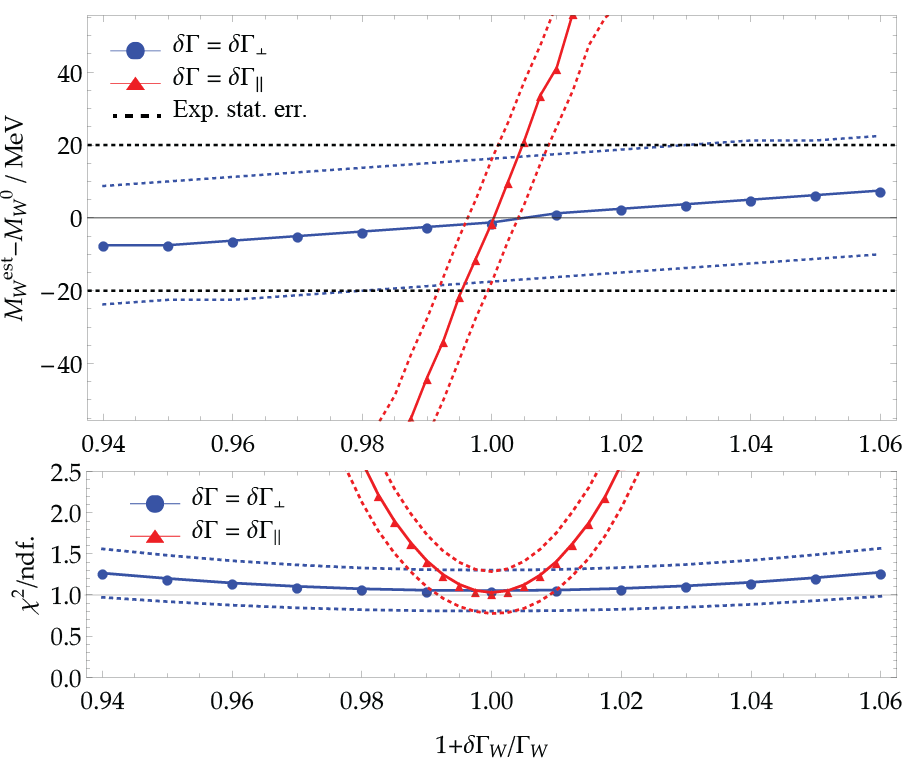}
\caption{The bias $m_W^{est}-m_W^0$ on the estimated $W$-mass relative to input mass in a fit to (a) the $m_T$-distribution and (b) the $p_{T\ell}$-distribtion, due to the presence of SMEFT operators. The SMEFT contribution is decomposed into $\delta \Gamma_{||}$ and $\delta \Gamma_{\perp}$ according to equations \eqref{eq:width_shift_decomp}-\eqref{eq:decomp_coefficient}. Note that this 1-d scan of the parameter space is only an approximation to a multi-dimensional parameter scan varying $\hat{m}_W, \hat{\Gamma}_W$ simultaneously. \label{fig:results}}
\end{figure*}

Using the same methodology, we have generated ``data samples'', keeping $\bar{m}_W=m_W^0$ but including the SMEFT variations discussed above. We obtain an estimated $\hat{m}_W^{est}$ by a binned log-likelihood fit of the simulated data samples to the template set, using the Poisson likelihood function
\begin{align*}
    -\ln \mathcal{L} = \sum - d_i \ln t_i + t_i + \ln[d_i !],
\end{align*} where $t_i$ is the expected event count in bin $i$ cf. the template, and $d_i$ is the data event count. We then investigate the measurement bias $\hat{m}_W^{est}- m_W^0$ as a function of the included SMEFT shift. We parametrize the shifts in $\delta \Gamma_{||}$ and $\delta \Gamma_\perp$ as described in Eqs.\eqref{eq:width_shift_decomp}-\eqref{eq:decomp_coefficient}, and vary both by $\pm 6 \%$, consistent with $\Lambda/\sqrt{C_i} \gtrsim 1$ TeV. The results are shown in Fig. \ref{fig:results}. We also show the $\chi^2$ goodness-of-fit parameters corresponding to the best $m_W$ fit, for each of the SMEFT shifted data samples. In order to obtain the plots, we have done the analysis for many different instantiations of Poissonian counting noise, and averaged the results. Only $\delta \Gamma_{||}$ yields a substantial bias on $\hat{m}_W^{est}$, but that this bias can be on the order of magnitude of $\sim 40 \, {\rm MeV}$ and still yield reasonable $\chi^2/n.d.f$ values. Note that for the plots shown $n.d.f = \{50,32\}$ for the $m_T, p_{T\ell}$ spectra respectively, so that a $\sim 2 \sigma$ shift in the goodness
of fit test is $|\chi^2/n.d.f| < \{1.4,1.5\}$ respectively yielding a bias $\hat{m}_W^{est}- m_W^0 \lesssim \{25,40\} \, {\rm MeV}$. However, this shift is due to the normalization change of the data, compared to the templates, as can be seen by the effect of $\delta \Gamma_\perp$ not introducing a significant bias. As the experimental collaborations float the overall normalization as a free parameter, this effect will not be present in the experimental result. Therefore, the measurement results are expected to reproduce $\bar m_W$ with no significant bias, when interpreted in the SMEFT. This is our main conclusion.
\paragraph{\bf VI. LEPII mass extractions.} At threshold we find the leading order (in $\beta$) SMEFT correction to the Born cross section approximation 
of $\sigma(e^+ \, e^- \rightarrow W^+ \, W^- \rightarrow S_i, S_j)$ is
\bea
\delta \frac{d \sigma}{d \Omega} &\simeq& \frac{\hat{\alpha}^2}{8 \, s  \, s_{\hat{\theta}}} \frac{\delta m_w^2}{\hat{m}_w^2}. 
\eea
Note the lack of s channel contributions due to possibly anomalous TGC
parameters in the threshold limit. These corrections first appear at order $\beta^3$, as is well known (see the review  \cite{Beenakker:1994vn}).
We agree with the discussion in Ref.\cite{Hagiwara:1987pb} that threshold extractions can be relatively insensitive to normalization corrections,
by fitting the shape of the cross section rise (in $\beta$) with a free normalization factor to account for the corrections in the first line. Further, only using specific final states
in such extractions, the remaining corrections of the form $\delta {\rm BR}$ can be corrected for. Unfortunately, the legacy LEP fit results on $\hat{m}_W$ using 
threshold data assumes the SM, and combines final states under the SM assumption. As such, a further theoretical error must be assigned when using this data for the neglect of SMEFT corrections. The requirement to use data away from threshold to reduce statistical errors in a manner that makes the measurement competitive with Tevatron results
introduces further SMEFT corrections due to anomalous TGC parameters. It is inconsistent to utilize the same data set for extractions of $\hat{m}_W,\hat{\Gamma}_W$
assuming vanishing anomalous TGC parameters in template fits at LEP, and then simultaneously use the same data with the extracted value of $\hat{m}_W,\hat{\Gamma}_W$
to constrain anomalous TGC parameters in the SMEFT. Considering the robustness of the Tevatron extractions, away from threshold LEP data should be reserved for direct constraints on anomalous TGC parameters. We discuss this issue in more detail in a companion paper \cite{fitpaper}.

\paragraph{\bf VII.  Conclusions.} In this paper we have examined if considering the SMEFT generalisation of the SM introduces a measurement bias in reported values of the $W^\pm$ mass in Tevatron data.
Such a bias would have to be incorporated as an extra component of the theoretical error assigned in using these measurements, and was potentially dominant over any pure theoretical error
associated with mapping these measurements to a bound on the Lagrangian parameters given by Eqn.~\ref{mwshift}. We have found\footnote{To our surprise.} that the effect of the SMEFT modification of the
measurement introduces a negligible bias in the extracted value of the $W^\pm$ mass. Our results show that mapping this very precise measurement to the SMEFT Lagrangian consistently (i.e. using Eqn.~\ref{mwshift}, a leading order result in the non-perturbative and perturbative expansion, to map to the SMEFT Lagrangian), is the dominant
issue in interpreting these results in the SMEFT. We stress that our results on the measurement bias are only the expected order of magnitude, however our estimate matches well with the estimate
given in Ref.\cite{Aaltonen:2013iut} on the correlation between errors in $\hat{\Gamma}_W$ and $\hat{m}_W$ assuming the SM. Two parameter extractions of  $\hat{m}_W,\hat{\Gamma}_W$, not assuming the SM, with the simultaneous reporting of a correlation matrix (expected to have small off diagonal entries \cite{troels}) using template fits to transverse variables
is very well motivated in the SMEFT and can be robustly interpreted. We encourage the experimental collaborations to also perform such an analysis on the legacy Tevatron data used to extract $\hat{m}_W$.

{\bf Acknowledgements} 
M.T. acknowledges generous support by the Villum Fonden and the Discovery center.
We thank L. Berthier for collaboration on the companion paper \cite{fitpaper} and comments on the manuscript. We thank members of the
Tevatron $W$ mass working group for helpful correspondence, and in particular Rafael Lopez de Sa. We also thank Andre Tinoco Mendes and Troels Petersen for very useful comments.
\vspace{-0.7cm}

\end{document}